\documentclass[twocolumn,twoside,slac_two]{revtex4}
\usepackage[dvipdfmx]{graphicx}
\usepackage{fancyhdr}
\usepackage{amsmath}
\begin{document}

\title{A Correlation Between Optical, X-ray, and Gamma-ray Variations in Blazar 3C 454.3}

%

\author{Yutaro Tachibana, Nobuyuki Kawai}
\affiliation{Department of Physics, Tokyo Institute of Technology, 2-12-1, Ohokayama, Tokyo, Japan, 152-8550}
\author{Sean Pike}
\affiliation{Depertment of Physics, Brown University, Providence, Rhode Island 02912, USA}
\author{on behalf of the MAXI Team and MITSuME Team}

\begin{abstract}
We present the light curve data of a remarkable blazer 3C 454.3 (z=0.859) in optical, X-ray, and gamma-ray bands. 
Since January 2008, we have been monitoring this object using the 50 cm MITSuME, a optical telescope, 
and detected several flares including extraordinary and simultaneous flares in the $\gamma$-ray and optical bands in 
November 2010. Additionally, the Monitor of All-sky Image (MAXI) has been observing 3C 454.3 continuously since 
August 2009.
Using these data and gamma-ray flux observed with Fermi-LAT, we discuss features and correlations of flux variations 
between the energy bands.
\end{abstract}

\maketitle

\thispagestyle{fancy}


\section{INTRODUCTION}
The flat-spectrum radio-loud quasar 3C 454.3 (z=0.859; Jackson and Brown, 1991) is a well known object 
as one of the most active and brightest sources in the gamma-ray sky. 
Optical and radio observations revealed that they are ascribed to strongly Doppler-boosted 
emission from relativistic electrons in a jet of plasma. 
The jet is oriented within $\sim$2$^{\circ}$ to the line of sight (Jorstad et al. 2005).  The ''small and big blue bump'' 
has been detected  by Raiteri et al. (2007) in the optical to ultraviolet band. 
The small blue bump is probably a mixture of a iron lines, Mg II lines, 
and Balmer continuum from the broad line region [Ogle et al. 2011]. 
On the other hand, the big blue bump is interpreted as a signature of the 
thermal accretion disk (eg. Pian et al. 1999). 
\begin{figure}
\centering
\includegraphics[width=82mm]{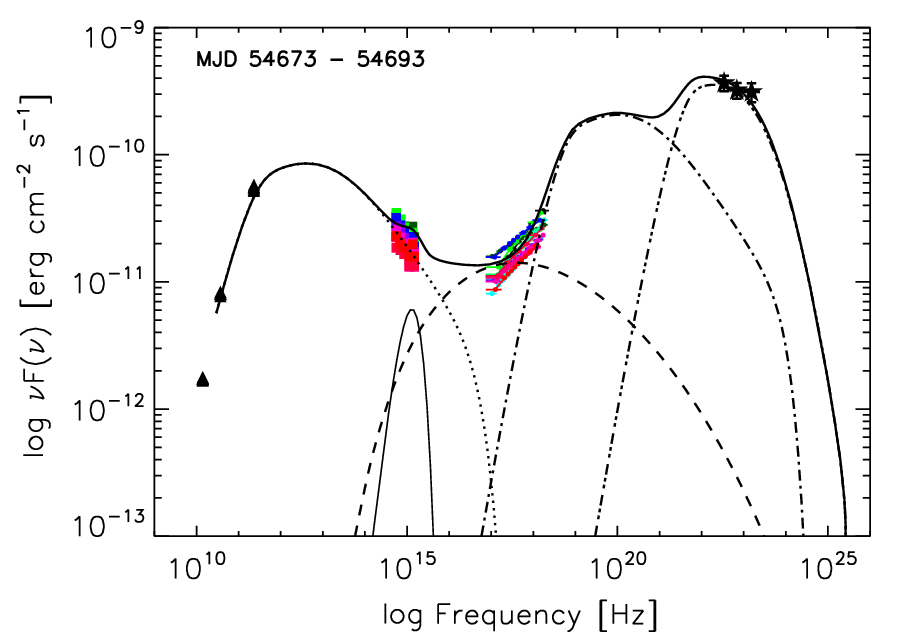}
\caption{The SED of 3C 454.3 during the period MJD = 54673-54693. 
The thin sold, dotted, dashed, dot-dashed, and the triple-dot-dashed lines represent 
the accretion disk component, the SSC component,  EC on the accretion disk, and EC on the broad line region (BLR), 
respectively (Vercellone et al. 2010). 
The sum of those the individual components is drawn by the thick solid line.}
\label{fig:SED}
\end{figure}
The X-to-$\gamma$-ray radiation from 3C 454.3 is commonly understood to be 
produced in the plasma jet through an Inverse-Compton (IC) scattering process 
off the same electrons that yield the radio to optical synchrotron emission. 
whereas, the origin of the seed photons for the IC scattering is not clear yet. 
Namely, (i) same synchrotron photons (SSC), 
(ii) photons coming from outside the jet (EC), 
or (iii) both of these would be possible to being seed photons. 
The X-ray range is most complicated because it might contain the high-energy 
tail of the synchrotron emission besides IC scattering component (Abdo et al. 2010). 
One example of the SED of 3C 454.3 is shown in Figure \ref{fig:SED} 
(quoted from Vercellone et al. 2010).

 Perhaps one of the most powerful tool to uncover the emission mechanism 
of the jet is time variability. Observing the with multi-wavelength for long period, 
and investigating correlations between different energy bands, 
we can build up a physical picture of the jet and its surroundings. 

 In this paper, we firstly show the light curve data in three energy bands, optical, 
X-ray, and $\gamma$-ray obtained by MITSuME and 
SMARTS\footnote[1]{http://www.astro.yale.edu/smarts/glast/home.php\#}, 
MAXI/GSC\footnote[2]{http://maxi.riken.jp/top/}, 
and Fermi/LAT\footnote[3]{http://fermi.gsfc.nasa.gov/ssc/data/access/lat/} 
respectively, secondly present some results of time domain analysis, 
and finally discuss about correlations of flux variations between these energy bands briefly.

\section{THE DATA AND ANALYSIS}

\subsection{Light Curves in Multi Wavelength}
\begin{figure}
\centering
\includegraphics[width=82mm]{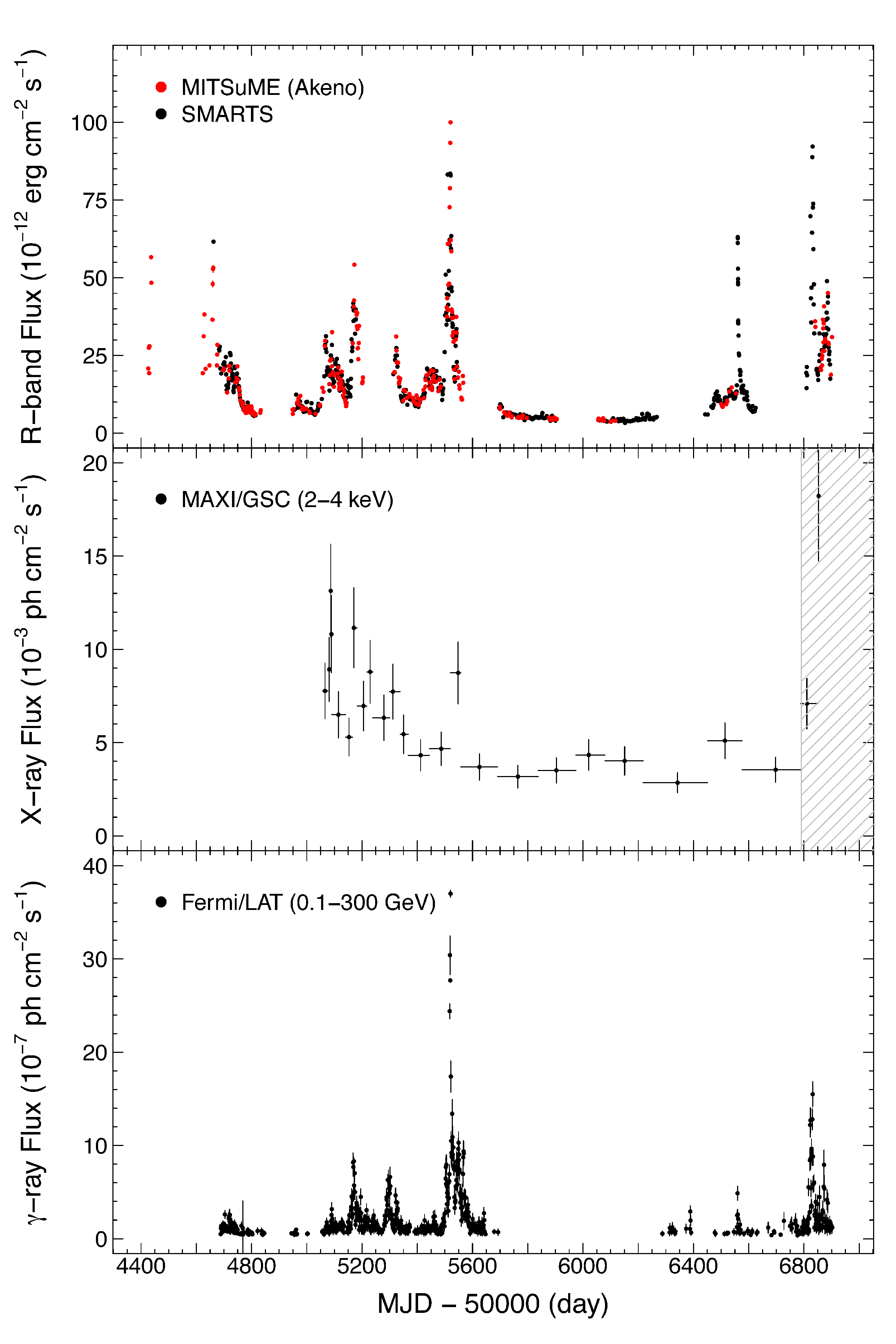}
\caption{The 7-years light curves of 3C 454.3 from MITSuME and SMARTS (R-band), 
X-ray (2-4 keV) flux from MAXI/GSC, $\gamma$-ray (0.1-300 GeV) flux from Fermi/LAT. 
In the shadowed area on the X-ray light curve, contaminating by a nearby X-ray source is suspected.}
\label{fig:LC}
\end{figure}
Figure \ref{fig:LC} shows the variation of R-band flux from MITSuME (Akeno Observatory, red points) 
and SMARTS (Bonning et al. 2012, black points),
 X-ray (2-4 keV) flux from MAXI/GSC, and $\gamma$-ray (0.1-300 GeV) daily averaged 
flux from Fermi/LAT. 
For converting the R-band magnitude to the flux, we assume that the 0 mag in R-band correspond to 
a flux of $1.42 \times 10^{-5} \rm {erg cm}^{-2} \rm{s}^{-1}$ (Sasada et al. 2011). 
 In the middle panel, the gray shaded region indicates the interval which may 
be contaminated by the nearby X-ray source IM Peg 
(see the ATel \#6296). 
To see the X-ray flux variations easily, we summed the daily flux data points
 until satisfying the following condition at each data points: Flux/Error $\geq$ 5. 
 
Variations, especially activities like flares, are seen to be completely 
correlated across the three energy bands. 
On the other hand, looking closely to variations in the light curves in the point of view of flare amplitudes, 
 the X-ray band seems to show lower variabilities
than the optical band and $\gamma$-ray bands. 
While it is certainly affected by a binning size,  it was also noticed in Vercellone et al. (2010).
Additionally, in the optical band and $\gamma$-ray band, flux responses originated from same events (same times) 
can be different from a flare to another.
We will present a detailed analysis on this point in section 2.4.

\subsection{Color-Magnitude Variation in Optical bands}
\begin{figure}[b]
\centering
\includegraphics[width=80mm]{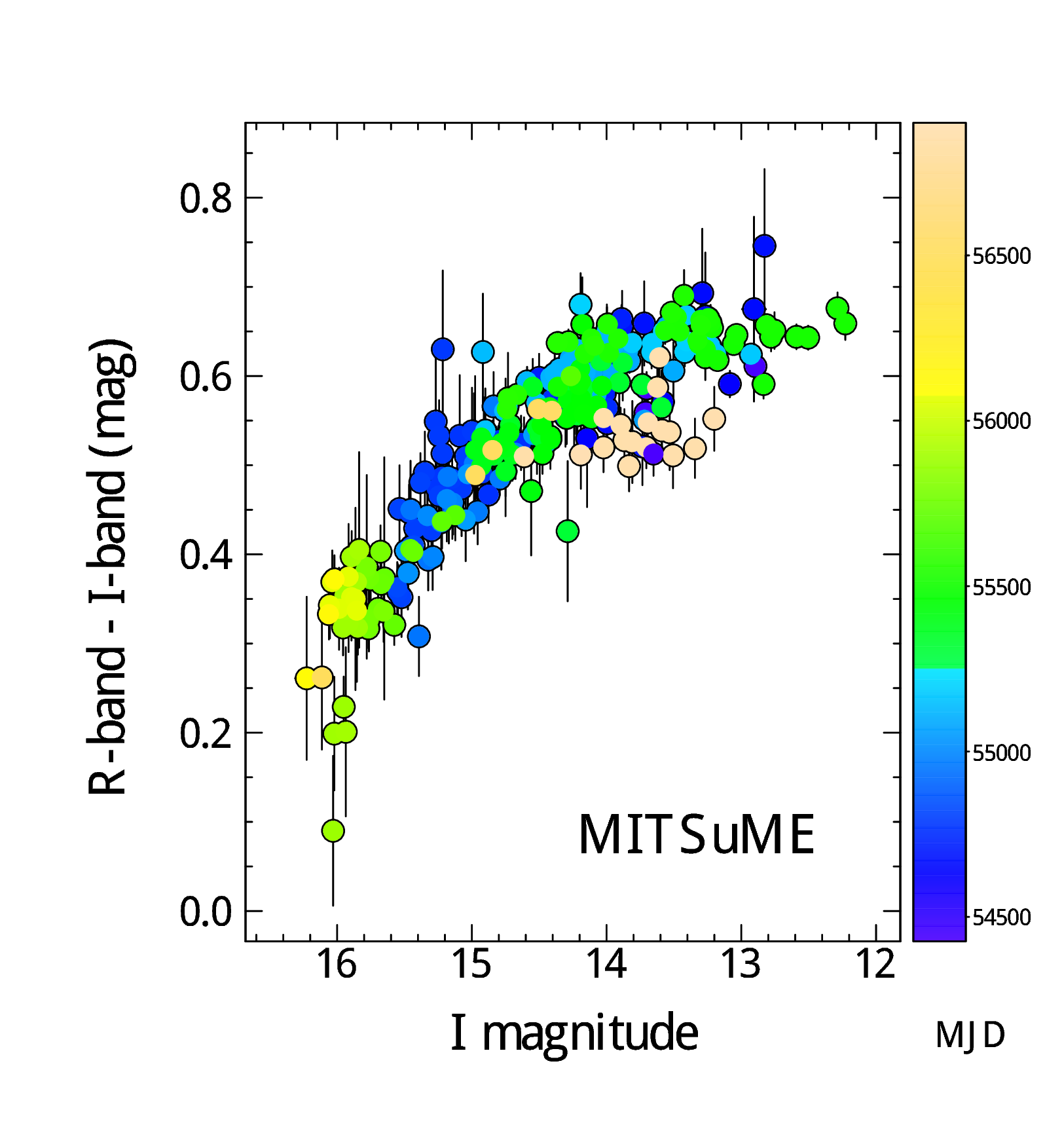}
\caption{R-I color vs. I-band magnitude for 3C 454.3. Colors indicate the date of the observation in units of MJD
as shown in the right bar. The redder-when-brighter trend and the plateau are shown in this digram. 
}
\label{fig:CMD}
\end{figure}

Color-magnitude analysis is useful for investigating the property of spectral behavior of the optical variation. 
We show the R-I color indices as a function of I-band magnitude in Figure \ref{fig:CMD}.
3C 454.3 shows flux fluctuation on a very short time scales (a few hours), so we used the R and I data taken 
simultaneously using the MITSuME tri-color camera for constructing the color magnitude diagram. 
As pointed out by Villata et al. (2006), this diagram shows the redder-when-brighter behavior 
which is interpreted as the contribution of a stable ''blue'' component, {\it i.e.} the thermal radiation 
from the accretion disk (big blue bump) lying under the variable ''red'' component 
{\it i.e.}  the synchrotron radiation from the relativistic electrons in the jet. 
When the jet emission dominates that of the accretion disk, 
a plateau from R-magnitude $\sim$14 towards the brightest end will emerge in the color-magnitude diagram.
It should be note that the magnitude of plateau have changed from $\sim$0.65 mag 
to $\sim$0.5 mag in a time between MJD $\sim$55500 and MJD $\sim$56800. 
This may indicate that some physical parameter of the accretion disk or the synchrotron emitting electrons 
(or both of these) are dramatically changed within this term.

\subsection{The Ratio of The Optical and The Gamma-ray Flux}
Studies about 3C 454.3 have revealed a close connection between the optical and $\gamma$-ray flux behavior 
not only in high activity states but also in non-flaring state (e.g. Bonning et al. 2008; Abdo et al. 2010a). 
Assuming photons in optical bands are produced by the synchrotron radiation of the relativistic electrons in the jet and 
$\gamma$-ray photons are produced through the IC scattering process by the same electrons, 
we can represent observed flux variations in these energy bands as 
\begin{align}
F_{\mathrm{opt}} &\sim N_e \delta^{3+\alpha_0} B^{1+\alpha_0} \\
F_{\gamma} &\sim N_e \delta^{4+2\alpha_g } U_{\mathrm{ext}}, 
\end{align}t
thus the flux ratio of the optical band and the $\gamma$-ray is predicted as following:
\begin{equation}
\frac{F_{\mathrm{opt}}}{F_{\gamma}} \sim 
B^{1+\alpha_0} U_{\mathrm{ext}}^{-1} \delta^{\alpha_0 -2\alpha_g -1},
\end{equation}
where $N_e$ is total number of emitting electron, $\delta$ is Doppler factor of an synchrotron emitting plasma, 
$\alpha_0$ and $\alpha_g$ are the spectral indices in the optical and $\gamma$-ray bands, respectively, 
$B$ is magnetic field in the emitting region, and $U_{\mathrm{ext}}$ is the external seed photon field 
(Chatterjee et al. 2012). 

\begin{figure}
\centering
\includegraphics[width=80mm]{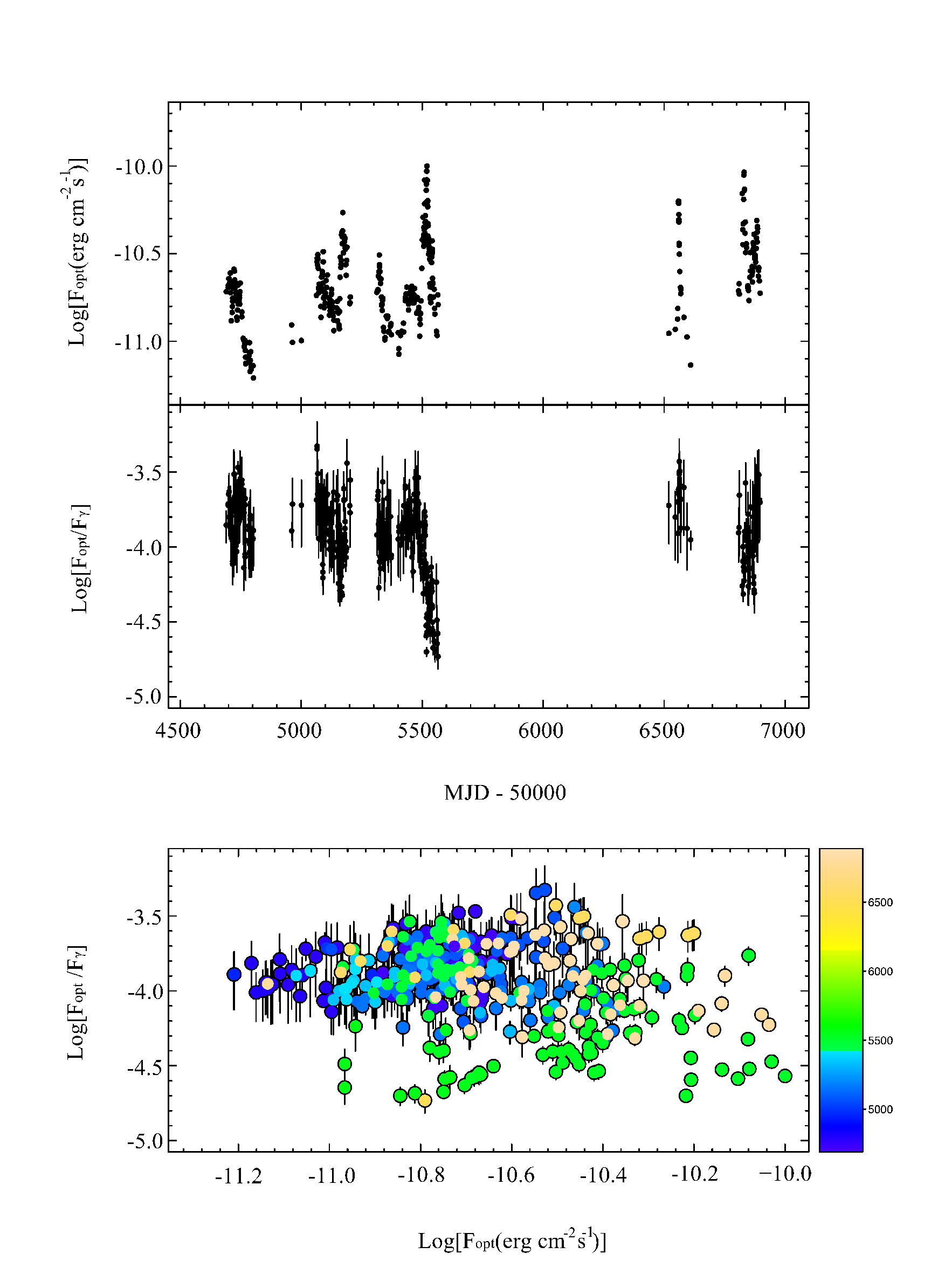}
\caption{The R-band light curve in the last 6 years observation with the corresponding ratio 
$F_{\mathrm{opt}}/F_{\gamma}$
as a function of time is shown in top and middle panel. 
In this analysis, we selected only the combination data of the R-band and the $\gamma$-ray 
to calculate the ratio between these energy bands, 
hence the number of data point is less than the original data (shown in Figure \ref{fig:LC}). 
Note that the unit of $F_{\mathrm{opt}}$ and $F_{\gamma}$ is [erg/cm$^{2}$/s] and 
[photons/cm$^{2}$/s], respectively, thus the ratio 
$F_{\mathrm{opt}}/F_{\gamma}$ is not a dimensionless quantity. 
The bottom panel is the ratio as a function of brightness level of the R-band and 
color indicates the date of the observation in units of MJD - 50000 as shown in the right bar.
}
\label{fig:ratio}
\end{figure}

Flux ratios calculated using the R-band flux and $\gamma$-ray flux pairs
observed on same days are shown in Figure \ref{fig:ratio}. 
From this figure,we can notice a trend that the flux ratio $F_{\mathrm{opt}}/F_{\gamma}$ 
becomes small as the source becomes bright. 
There seems to be some time lag in this tendency 
(the decrease of the ratio lags the increase of the brightness), 
and this point will be investigated in section 2.3.1.
If the variation is only due to a change in $N_e$, 
the ratio $F_{\mathrm{opt}}/F_{\gamma}$ will be constant 
because the ratio does not depend on $N_e$, as shown in Eq.(3) . 
In addition, as can be appreciated from Eq.(1) and Eq.(2), 
if the variation arose from a change in $B$ or $U_{\mathrm{ext}}$, 
the optical and the $\gamma$-ray variations will not be correlated. 
Therefore, this supposition is inconsistent with the previous research results, except the so called 
''$\gamma$-ray orphan'' optical-UV flare (eg. Vercellone et al. 2011). 
For these reason, this trend may indicates that the Doppler factor $\delta$ plays a big role 
in the flux fluctuation in the optical and  $\gamma$-ray bands. 
The origin of flux variation will also be discussed in section 2.5.

\subsubsection{Z-transformed Discrete Cross Correlation Function analysis}
\begin{figure}
\centering
\includegraphics[width=90mm]{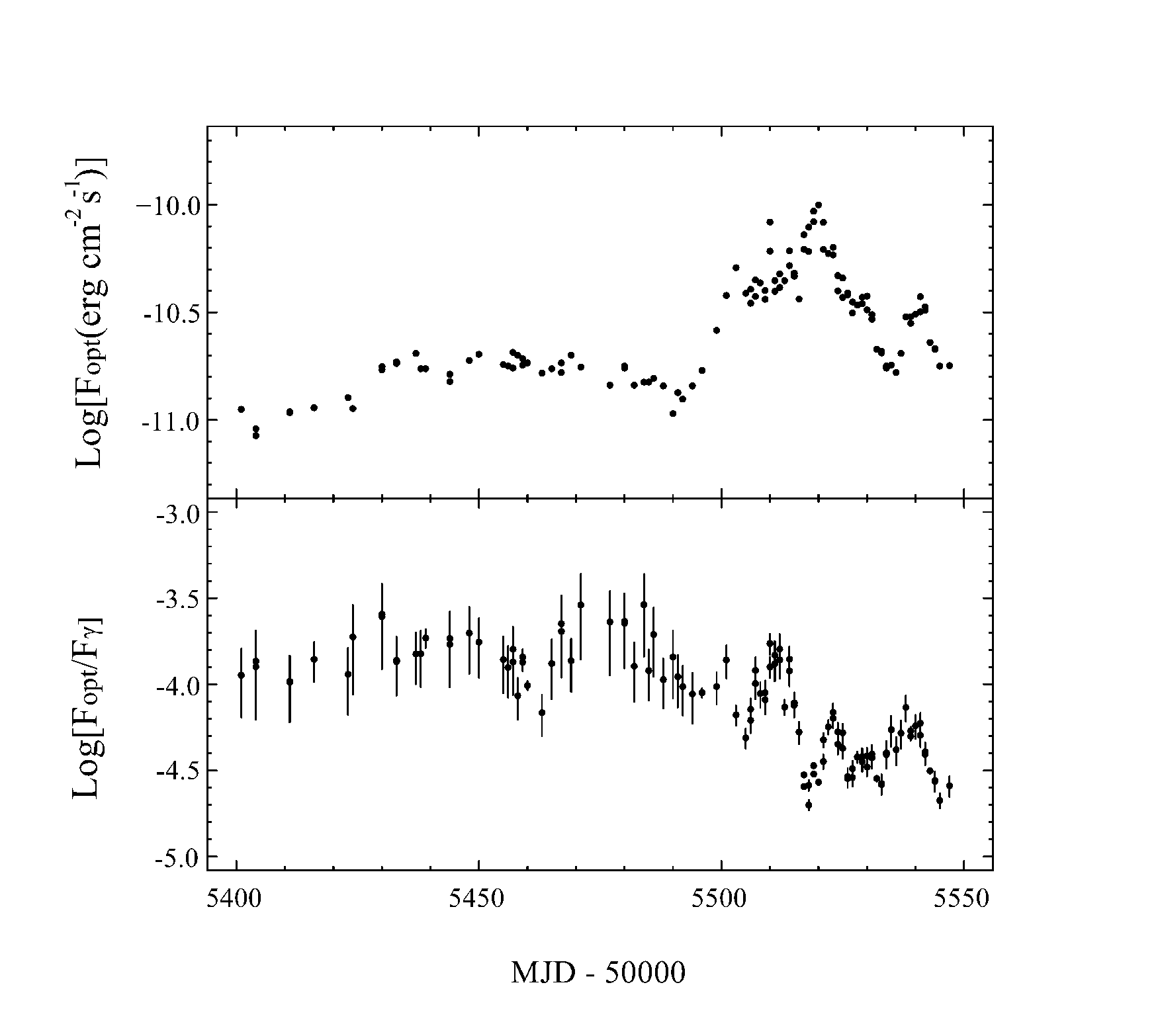}
\includegraphics[width=75mm]{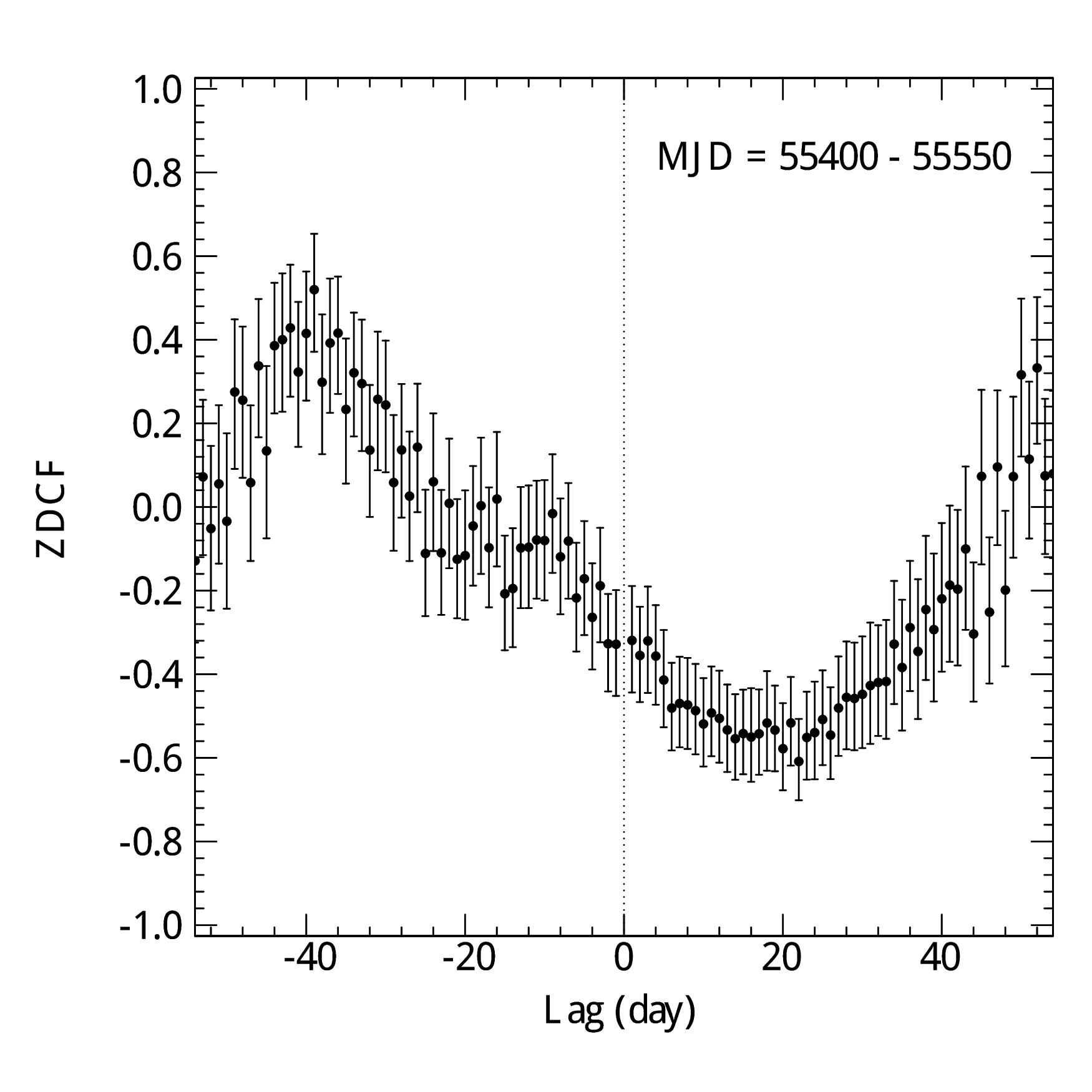}\hspace*{0em} 
\caption{The R-band light curve in MJD = 55400-55550 with the corresponding ratio 
$F_{\mathrm{opt}}/F_{\gamma}$
as a function of time is shown in top and middle panel. 
The bottom panel is the ZDCF of  the flux ratio $F_{\mathrm{opt}}/F_{\gamma}$ and R-band light curve 
during that interval. 
The time delay is defined as positive if the flux ratio variations lead that at R-band flux.
}
\label{fig:ZDCF}
\end{figure}

To investigate the time lag between the flux ratio $F_{\mathrm{opt}}/F_{\gamma}$ and 
the brightness of the optical band, we employed the Z-transformed Discrete Cross Correlation Function 
(ZDCF) introduced by Alexander (1997). 
This method can correct several biases of the discrete correlation function of Edelson \& Krolik (1988) 
by using Fisher's z-transform and equal population binning (see the original paper for more detail). 
The calculated ZDCF between the flux ratio $F_{\mathrm{opt}}/F_{\gamma}$ versus 
the optical band flux $F_{\mathrm{opt}}$ during MJD = 55400 -- 55550 is shown in Figure \ref{fig:ZDCF}. 
In this period the most prominent change of the ratio and brightness was seen as shown in Figure \ref{fig:ratio}.
A positive time lag indicate that the variation of  the flux ratio is delayed with respect to that of the brightness, 
thus the decrease of the flux ratio lags by typically about 20 day with respect to the increase of the flux in the optical band. 
The maximum likelihood 1$\sigma$ error estimate calculated for the points between time-lags = $-$40 to 40 days is 
22.00 $^{+4.12}_{-7.49}$ days. 
The physical meaning of this value is not clear at present.
In a qualitative manner, however, we find the behavior of the flux ratio has the following two features:
(i) fluctuations on relatively short time-scales (a few days) coincides with flare like activities, 
and (ii) a relatively long time-scale decreasing trend in active phase (MJD $\gtrsim$ 55490) 
in the middle panel of Figure \ref{fig:ratio}. 
The 22.00 day lag found above is probably related to the latter feature, the long-term decreasing trend of the flux ratio associated with the large swings of the optical flux.

For the behavior of the flux ratio, described above as (i) and (ii), we can interpret this by a existence of parameters 
which change on long time scales ($>$50 days) besides the Doppler factor $\delta$. 
Namely, if the external seed photon field $U_{\mathrm{ext}}$ increase or 
if magnetic field in a emitting region decreases gradually, 
the flux ratio will turn downward as shown in Figure \ref{fig:ratio} according the changes in these parameters.
In addition, it is also possible that the baseline of the Doppler factor progressively increases as
 an angle between the line of sight and the direction of the jet progressively decreases. 

\subsection{flare analysis}
\begin{figure}[t]
\centering
\includegraphics[width=70mm]{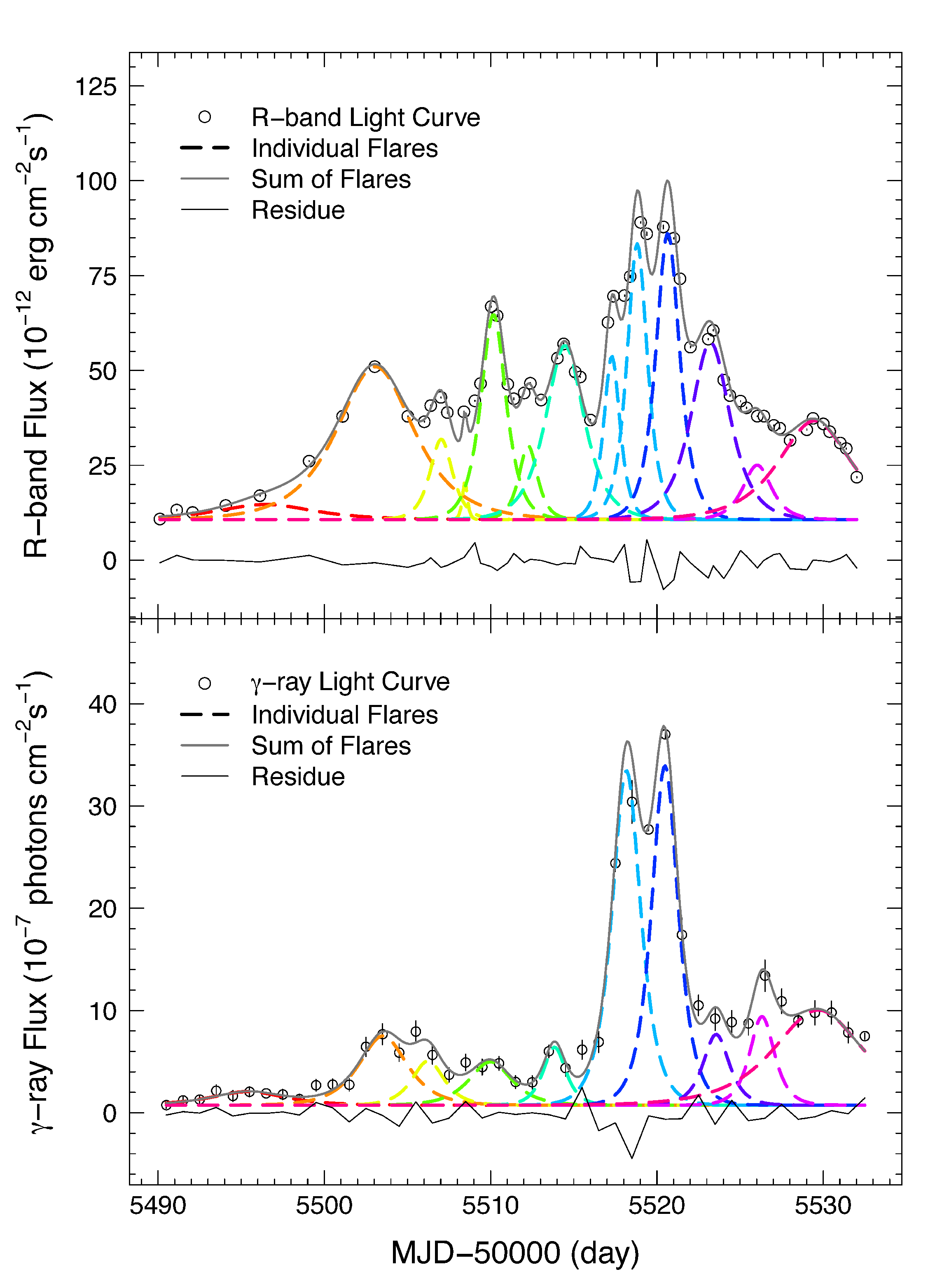}
\caption{The decomposed light curves of R-band and $\gamma$-ray into individual flares are drawn by 
the dotted curve. 
The gray solid curve corresponds to summed flux of the modeling light curves, 
black open circles dente the observed light curve, 
the gray dotted line shows a residual flux, 
and same colors indicate that these flares were yielded by a same event 
in each panel. 
}
\label{fig:flare}
\end{figure}
The fluctuations in the optical and $\gamma$-ray light curves of blazars can be interpreted as a superposition of 
individual flares caused by same kind of events lying on a steady baseline flux 
(e.g. Abdo et al 2010c, Chattergee et al. 2012). 
For the optical band, for example, the steady component and the flaring component can be interpreted as 
thermal radiation from the accretion disk and synchrotron radiation in the jet, respectively.
In order to investigate properties of the variable component, we decompose the light curves 
into individual flares represented by the following function:
\begin{equation}
F(t) = F_c + F_0
\left(
e^{\frac{t_0 - t}{T_r}} + 
e^{- \frac{t_0 - t}{T_d}}
\right) ^{-1},
\end{equation}
where $F_c$ is an assumed constant flux level underlying the flare, $F_0$ is the amplitude of the flare, 
$t_0$ is the epoch of the peak of the flare, and $T_r$ and $T_d$ are the rise and decay time of the flare. 
$F_0$ was constrained to be equal to the lowest value of the flux during the focusing interval. 
In our analysis, we assume $T_r = T_d$ because the acceleration and cooling time scales of the relativistic electrons 
are expected to be substantially shorter than the light-crossing timescale of the synchrotron emitting region. 
Figure \ref{fig:flare} shows the optical and $\gamma$-ray light curves for the period containing the brightest $\gamma$-ray flare 
 in the Fermi observation from 3C 454.3.
The individual flares presented by dashed lines of same colors in the R-band and the $\gamma$-ray light curves 
indicate that these variations were produced by the same origin.
One can see that almost all flares are correlated in this period but flux responses in these energy bands 
originated from same events may be very different.
On the other hand, the duration times of these flares are roughly equal. 
These aspects might be representing the origin of the observed flares. 
Assuming that the sub-equal duration time of the optical and $\gamma$-ray flares is caused by 
a light-travel time in a same emitting region and 
the difference of flux response is made by a variance of physical parameters in the emitting region 
such as $\delta$ or $B$, 
each flare observed might be yielded by different blobs of denser plasma possessing almost same size 
with substantially different parameters.

\subsection{Gamma-ray/Optical Correlation}
As shown in Eq.(1) and (2), the optical flux $F_{\mathrm{opt}}$ and the $\gamma$-ray flux $F_{\gamma}$ 
are related to each other through several parameters. 
For example, if the flux variation is due to a change in total number of emitting electron $N_e$, 
$F_{\gamma} \propto F_{\mathrm{opt}}$, and if it is due to a change in the Doppler factor $\delta$, 
$F_{\gamma} \propto F_{\mathrm{opt}}^{(4+2\alpha_g)/(3+\alpha_0)}$. 
\begin{figure}[t]
\centering
\includegraphics[width=70mm]{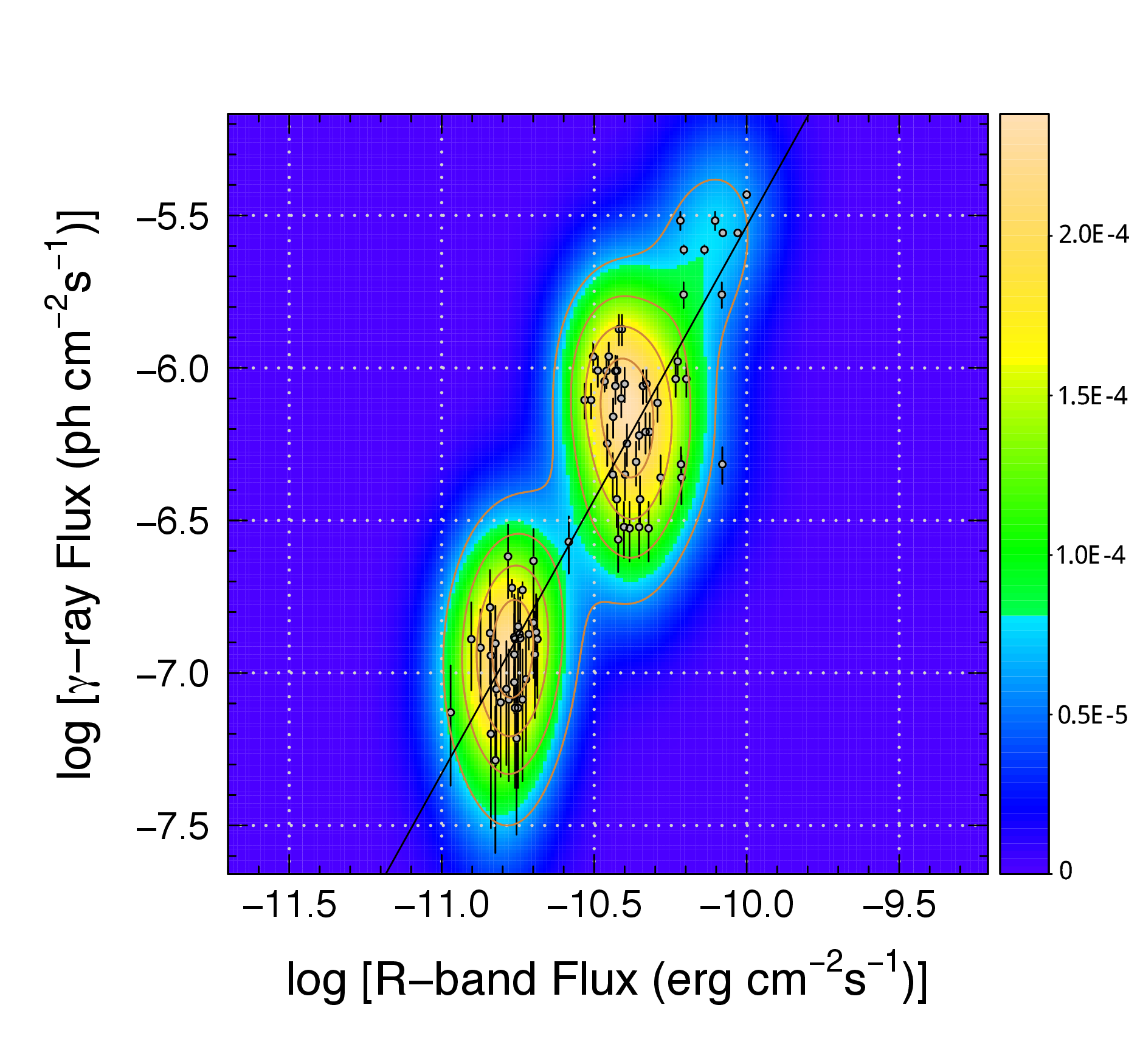}
\includegraphics[width=70mm]{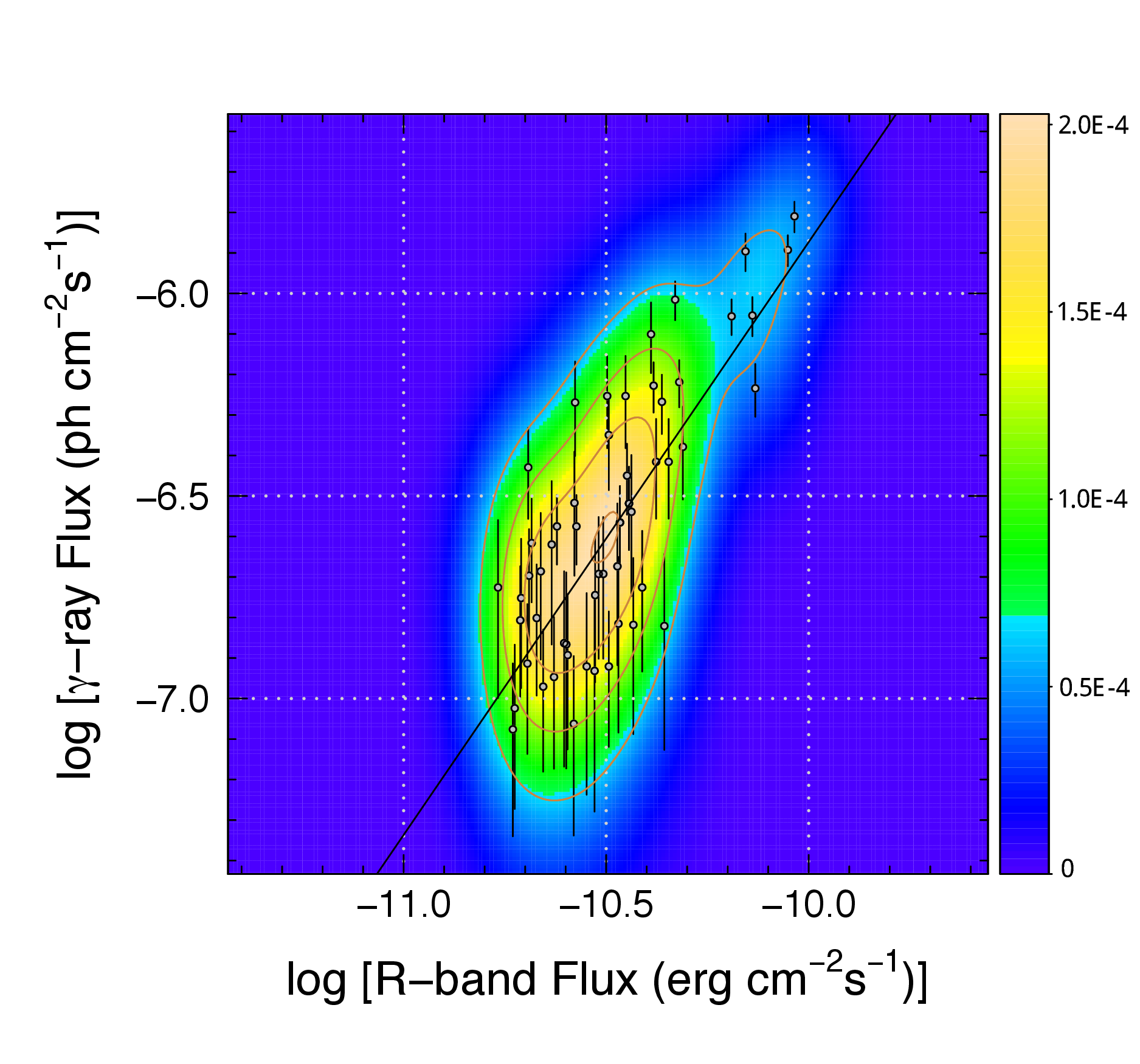}
\caption{Log$(F_{\mathrm{opt}})$ - Log$(F_{\gamma})$ dependences 
durring MJD = 55434-55532 and MJD = 56800-56910. 
Color contours indicate the density of data points and the solid line denote the numerical relation 
between the fluxes in those two energy bands.
}
\label{fig:cor}
\end{figure}
To investigate this relationship, we calculate the slope of the distribution in the Log$(F_{\mathrm{opt}})$ vs. Log$(F_{\gamma})$ space. 
Data points in the Figure \ref{fig:cor} show the observed $\gamma$-ray and the optical fluxes for the
two substantially bright periods: MJD = 55434 - 55532 (upper panel), and 55680 - 56910 (bottom panel). 
Color contours indicate the density of data points and 
the lines denote the regression relation between the fluxes in those two bands, 
e.g., $F_{\gamma} \propto F_{\mathrm{opt}}^X$, 
where $X$ is the slope of  Log$(F_{\mathrm{opt}})$ vs. Log$(F_{\gamma})$. 
In MJD = 55434 - 55532 and MJD = 56800 - 56910, parameter X is calculated as 1.79 and 1.46, respectively. 
According to Ackermann et al. (2010), in MJD = 55060 - 55160, OIR spectral index 
from SMARTS light curve is 1.55 $\pm$ 0.05, 
and the $\gamma$-ray spectral index is 1.5 $\pm$ 0.1. 
For $\alpha_0 = 1.55$ and $\alpha_g = 1.5$, we can obtain the relation $F_{\gamma} \propto F_{\mathrm{opt}}^{1.54}$.
Thus, if we assume that the spectral index was largely unchanged and the $\gamma$-rays are produced by EC processes, 
this result seems to imply that the variation is mainly due to a change in $\delta$ 
and support the suggestion in section 2.3. 

\section{SUMMARY AND CONCLUSIONS}
In this paper, we present the long-term light curve data of 3C 454.3 in three energy bands 
{\it i.e.} optical (R-band), X-ray (2-4 keV), and $\gamma$-ray (0.1-300 GeV) provided by 
MITSuME and SMARTS, MAXI/GSC, and Fermi/LAT, and reveal several time variability properties 
using the optical and the $\gamma$-ray flux data. 
Our main conclusions are as follows.
\begin{enumerate}
\item{
Using R-band and I-band data from MITSuME, the redder-when-brighter behavior and 
the ''saturation'' effect from R-magnitude $\sim$14 towards the brightest end are confirmed by the color-magnitude diagram. 
These results were pointed out by some previous research ({\it e.g.} Raiteri et al. 2008, Zhai et al. 2011). 
The remarkable result in this research is the finding of a sign of change in the plateau magnitude. 
The physical parameter of the accretion disk or the relativistic electron in the jet might have changed gradually and significantly.
}
\item{
We investigate a relation between the optical-band flux $F_{\mathrm{opt}}$ 
and the flux ratio of it to the $\gamma$-ray flux $F_{\mathrm{opt}}/F_{\gamma}$. 
The flux ratio becomes decreasing as the source becomes brightening, 
and this might indicate that the Doppler factor of an synchrotron emitting plasma is playing the most important role 
in the variations in the optical and $\gamma$-ray flux from 3C 454.3. 
More detailed analysis is performed for the period of MJD = 55400 to 55550. 
During this period, the flux ratio shows the interesting behaviors, namely, 
(i) steep dropping corresponding with the flares of the optical and the $\gamma$-ray, and 
(ii) slow declining during the highly active phase (MJD $\gtrsim$ 55490). 
It might be interpreted by an existence of a change of physical parameters in the jet besides the Doppler factor $\delta$ 
or increasing of baseline of the Doppler factor gradually for some reasons.
In addition, the $\sim$ 20 days time-lag between the flux ratio and the optical flux variation is detected. 
A Interpretation of this phenomenon will be addressed in a future paper. 
}
\item{
We performed a decomposition of the optical and $\gamma$-ray light curves of 3C 454.3 
during its large flaring activity (MJD = 55490-55540). 
Almost all flares are correlated in this period, though flux responses vary significantly from one event to the other. 
This might indicate that the size of plasma blobs which radiate the optical and the $\gamma$-ray photons are almost same 
while those parameters besides its size are diverse.
}
\item{
For the two bright periods of 3C 454.3, we investigate the numerical relationship 
between $F_{\mathrm{opt}}$ and $F_{\gamma}$, 
and $F_{\gamma} \propto F_{\mathrm{opt}}^{1.79}$ and
$F_{\gamma} \propto F_{\mathrm{opt}}^{1.46}$ are obtained
in MJD = 55434 - 55532 and MJD = 56800 - 56910, respectively. 
These are roughly consistent with 
$F_{\gamma} \propto F_{\mathrm{opt}}^{1.54}$ 
which is expected if the variation is due to a change in the Doppler factor $\delta$.
}
\end{enumerate}

\bigskip 

\begin{thebibliography}{9}   

\bibitem{ab}
Abdo et al., ApJ716 (2010) 835-849.
\bibitem{ac}
Ackermann et al., ApJ721 (2010) 1383-1396.
\bibitem{al}
Alexander, MNRAS285 (1997) 891-897.
\bibitem{b}
Bonning et al., ApJ756 (2012) 13-28.
\bibitem{c}
Chatterjee et al., ApJ749 (2012) 191-203.
\bibitem{c}
Jackson and Brown, MNRAS250 (1991) 422-431.
\bibitem{j}
Jorstad et al., AJ130 (2005) 1418-1465.
\bibitem{ma1}
Matsuoka et al., PASJ61 (20059) 999-1010.
\bibitem{ma2}
Marshall et al., ApJS156 (2005) 13-33.
\bibitem{mi}
Miniutti et al., PASJ59 (2007) 315-325.
\bibitem{na}
Nandra et al., ApJ476 (1997) 70-82.
\bibitem{no}
Noda et al., ApJ771 (2013) 100-112.
\bibitem{o}
Ogle et al., ApJS195 (2011) 19-38.
\bibitem{p}
Pian et al., ApJ521 (1999) 112-120.
\bibitem{r1}
Raiteri et al., A\&A473 (2007) 819-827.
\bibitem{r2}
Raiteri et al., A\&A534 (2011) 87-102.
\bibitem{s}
SaSada et al., PASJ63 (2011) 489-497.
\bibitem{v1}
Vercellone et al., ApJ712 (2010) 405-420.
\bibitem{v2}
Vercellone et al., ApJ736 (2011) L38-L45.
\bibitem{v3}
Villata et al. A\&A453 (2006) 817–822
\bibitem{z}
Zhou et al., NewA36 (2015) 19-25.
\bibitem{fermi}
http://fermi.gsfc.nasa.gov/ssc/data/access/lat/
\bibitem{maxi}
http://maxi.riken.jp/top/
\bibitem{smart}
http://www.astro.yale.edu/smarts/glast/home.php\#

\end{thebibliography}

\end{document}